\documentclass[letterpaper, 10 pt, conference]{ieeeconf}  %

\IEEEoverridecommandlockouts                              %
\usepackage{amsmath,amssymb,amsfonts,mathrsfs,amsthm}
\usepackage{graphicx} %
\usepackage{xcolor}
\usepackage{adjustbox}
\usepackage{pifont}%
\usepackage{tabularray}
\usepackage{flushend}
\usepackage[utf8]{inputenc}
\usepackage[T1]{fontenc}
\usepackage[lighttt]{lmodern}
\usepackage{listings}

\newcommand{\codestyle}{\footnotesize\ttfamily}
\newcommand{\codestylenormal}{\small\ttfamily}

\lstnewenvironment{python}
{\lstset{
    sensitive=true,
    language=python,
    frame=lines,
    xleftmargin=\parindent,
    belowcaptionskip=1\baselineskip,
    basicstyle=\codestyle, 
    keywordstyle=\footnotesize\bfseries\ttfamily, 
    commentstyle=\footnotesize\slshape\ttfamily, 
    stringstyle=\footnotesize\ttfamily, 
    numberstyle=\footnotesize\ttfamily, 
    breaklines=true,
    showstringspaces=false,
    showspaces=false,                
    showtabs=false,                  
    tabsize=2,
    breakatwhitespace=false,         
    breaklines=true,                 
    captionpos=b,                    
    keepspaces=false,                 
    numbers=left,                    
    numbersep=5pt,                  
}}{}

\lstnewenvironment{pythonoutput}
{\lstset{
    frame=lines,
    basicstyle=\scriptsize\slshape\ttfamily, 
    tabsize=2,
}}{}

\usepackage{cite}
\usepackage[hidelinks]{hyperref} 

\SetTblrInner{rowsep=1pt}
\newcommand{\cmark}{\ding{51}}%

\title{pycvxset: A Python package for convex set manipulation}
\author{Abraham P. Vinod
\thanks{A.~P.~Vinod is with Mitsubishi Electric Research Laboratories, Cambridge, MA 02139, USA. Email: {\tt\footnotesize \{abraham.p.vinod\}@ieee.org}.}%
}
\date{}

\begin{document}
\newcommand{\bbc}{\mathbb{C}}
\newcommand{\bbe}{\mathbb{E}}
\newcommand{\bbn}{\mathbb{N}}
\newcommand{\bbp}{\mathbb{P}}
\newcommand{\bbr}{\mathbb{R}}
\newcommand{\bbs}{\mathbb{S}}
\newcommand{\bbz}{\mathbb{Z}}
\newcommand{\bfc}{\mathbf{C}}
\newcommand{\bfx}{\mathbf{X}}
\newcommand{\cala}{\mathcal{A}}
\newcommand{\calb}{\mathcal{B}}
\newcommand{\calc}{\mathcal{C}}
\newcommand{\cald}{\mathcal{D}}
\newcommand{\cale}{\mathcal{E}}
\newcommand{\calf}{\mathcal{F}}
\newcommand{\calg}{\mathcal{G}}
\newcommand{\calh}{\mathcal{H}}
\newcommand{\cali}{\mathcal{I}}
\newcommand{\calj}{\mathcal{J}}
\newcommand{\calk}{\mathcal{K}}
\newcommand{\call}{\mathcal{L}}
\newcommand{\calm}{\mathcal{M}}
\newcommand{\caln}{\mathcal{N}}
\newcommand{\calo}{\mathcal{O}}
\newcommand{\calp}{\mathcal{P}}
\newcommand{\calq}{\mathcal{Q}}
\newcommand{\calr}{\mathcal{R}}
\newcommand{\cals}{\mathcal{S}}
\newcommand{\calt}{\mathcal{T}}
\newcommand{\calu}{\mathcal{U}}
\newcommand{\calv}{\mathcal{V}}
\newcommand{\calw}{\mathcal{W}}
\newcommand{\calx}{\mathcal{X}}
\newcommand{\caly}{\mathcal{Y}}
\newcommand{\calz}{\mathcal{Z}}
\newcommand{\scra}{\mathscr{A}}
\newcommand{\scrb}{\mathscr{B}}
\newcommand{\scrc}{\mathscr{C}}
\newcommand{\scre}{\mathscr{E}}
\newcommand{\scrg}{\mathscr{G}}
\newcommand{\scrh}{\mathscr{H}}
\newcommand{\scrl}{\mathscr{L}}
\newcommand{\scro}{\mathscr{O}}
\newcommand{\scrp}{\mathscr{P}}
\newcommand{\scrr}{\mathscr{R}}
\newcommand{\scrs}{\mathscr{S}}
\newcommand{\scru}{\mathscr{U}}
\newcommand{\scrv}{\mathscr{V}}
\newcommand{\scrw}{\mathscr{W}}
\newcommand{\scrx}{\mathscr{X}}

\newcommand{\Nint}[2]{\mathbb{N}_{[#1, #2]}}

\newcommand{\changepolytopetocz}{\dagger}
\newcommand{\pycvxset}{\texttt{pycvxset}}
\newcommand{\cvxpy}{\texttt{CVXPY}}
\newcommand{\pycddlib}{\texttt{pycddlib}}
\newcommand{\scipy}{\texttt{scipy}}
\newcommand{\matplotlib}{\texttt{matplotlib}}
\newcommand{\gurobi}{\texttt{GUROBI}}
\newcommand{\polytope}{\texttt{Polytope}}
\newcommand{\conzono}{\texttt{ConstrainedZonotope}}
\newcommand{\ellipsoid}{\texttt{Ellipsoid}}

\maketitle

\begin{abstract}
This paper introduces pycvxset, a new Python package to manipulate and visualize convex sets. We support polytopes and ellipsoids, and provide user-friendly methods to perform a variety of set operations. For polytopes, pycvxset supports the standard halfspace/vertex representation as well as the constrained zonotope representation. The main advantage of constrained zonotope representations over standard halfspace/vertex representations is that constrained zonotopes admit closed-form expressions for several set operations. pycvxset uses CVXPY to solve various convex programs arising in set operations, and uses pycddlib to perform vertex-halfspace enumeration. We demonstrate the use of pycvxset in analyzing and controlling dynamical systems in Python. pycvxset is available at \href{https://github.com/merlresearch/pycvxset}{https://github.com/merlresearch/pycvxset} under the AGPL-3.0-or-later license, along with documentation and examples. 
\end{abstract}
\section{Introduction}

Set-based methods provide a formal framework to analyze and control dynamical systems.
Such methods are often used in set propagation and reachability analysis where the goal is to characterize system states and a family of controllers with some desirable properties~\cite{blanchini2008set,bertsekas_1971,borrelli2017predictive,althoff2021set}.
For example, in spacecraft rendezvous, we can use set-based methods to define a range of acceptable positions and velocities along the nominal spacecraft trajectory to ensure a safe abort when needed~\cite{vinod2021abort,vinod2024projection,vinodCDC2024}.
See~\cite{borrelli2017predictive,blanchini2008set,scott_constrained_2016,raghuraman2022set,yang2022efficient} for other applications of set-based control methods.

For linear systems, set-based methods yield practical implementations using efficient set representations like ellipsoids and polytopes.
However, several set operations are not closed in ellipsoids~\cite{Elltoolbox}, and certain set operations in the standard vertex/halfspace representation of polytopes require computationally expensive vertex-halfspace enumeration~\cite{borrelli2017predictive}.
Recently, constrained zonotopes have been proposed for performing exact set operations on polytopes, since 1) they provide an equivalent representation of polytopes, and 2) they allow for closed-form expressions for several set operations~\cite{scott_constrained_2016,raghuraman2022set,vinod2024projection,vinodCDC2024,yang2022efficient}.

Several open-source software toolboxes implement some or all of these set representations and their operations in various languages~\cite{pycddlib,pytope,polytope,pypoman,althoff2015introduction,MPT3,zonolab,Elltoolbox}.
Together, these toolboxes have been instrumental in improving the access to set-based methods for reachability and trajectory optimization for the broader dynamical systems and control community.
While existing tools in Python~\cite{pycddlib,pytope,polytope,pypoman} have primarily focused on polytopes, MATLAB tools~\cite{althoff2015introduction,MPT3,zonolab,Elltoolbox} are more mature and accommodate more set representations.
pycvxset aims to help bridge this gap.

This paper \emph{introduces} \pycvxset{}, \emph{a Python package to manipulate and visualize convex sets}.
With \pycvxset{}, we hope to bring the recent progress made in set representations especially constrained zonotopes~\cite{vinodCDC2024,scott_constrained_2016,raghuraman2022set,vinod2024projection} to Python. 
\pycvxset{} extends \texttt{pytope}~\cite{pytope} to include ellipsoidal and constrained zonotopic set representations, broaden the capabilities of \polytope{} class including 3D plotting, and integrates with \cvxpy{} for use in constrained control. \pycvxset{} is extensively tested and documented for reliability and ease of use.

\section{Set representations and operations}

\subsection{Set representations}
We briefly review the various set definitions supported by \pycvxset{}. 
We say two set representations are \emph{equivalent} when the sets they represent contain each other.
Note that \pycvxset{} only supports bounded sets.

From~\cite{borrelli2017predictive,scott_constrained_2016}, the following three set representations are equivalent:
\begingroup
    \makeatletter\def\f@size{8.5}\check@mathfonts
\begin{align}
    \calp(V) &= \left\{x\in\bbr^n\ \middle|\begin{array}{c}
        \exists \theta\in\bbr^{N},\ x=V^\top\theta \\
         1^\top \theta = 1,\ \theta \geq 0
    \end{array}\right\}\label{eq:vrep},\\
    \calp(A,b,A_e,b_e) &= \{x\in\bbr^n\ |\ Ax \leq b,\  A_e x = b_e\}\label{eq:hrep},\\
    \calp(G,c,A_e,b_e) &=\left\{x\in\bbr^n\middle|\begin{array}{c}
        \exists \xi\in\bbr^N,\ x=G\xi + c,\\
        \|\xi\|_\infty \leq 1,\ A_ex=b_e
    \end{array}\right\}\label{eq:cz},
\end{align}
\endgroup
with appropriate dimensions for $V,A,b,A_e,b_e,G,c$.
Specifically, \eqref{eq:vrep} is the vertex representation, \eqref{eq:hrep} is the halfspace representation, and \eqref{eq:cz} is the constrained zonotopic representation of a polytope.
While the equivalence of \eqref{eq:vrep} and \eqref{eq:hrep} is well-known~\cite{borrelli2017predictive}, the equivalence to a constrained zonotope representation was recently established in~\cite[Thm. 1]{scott_constrained_2016}. 

The following sets are special cases of polytopes:
\begingroup
    \makeatletter\def\f@size{8.5}\check@mathfonts
\begin{align}
    \calr(l, u) &= \{x\in\bbr^n\ |\ l\leq x \leq u\}\label{eq:rect}\\
    \calr(c, h) &= \{x\in\bbr^n\ |\ \forall i\in\{1,2,\ldots,n\},\ |x_i - c_i|\leq h_i\}\label{eq:centered_rect}\\
    \calz(G,c) &=\{x\in\bbr^n\ |\ \exists \xi\in\bbr^N,\ x=G\xi + c,\ \|\xi\|_\infty \leq 1\}\label{eq:z}.
\end{align}
\endgroup
for finite vectors $l,u,c,h\in\bbr^n$ and appropriately dimensioned matrix $G$.
Here, \eqref{eq:rect} and \eqref{eq:centered_rect} represent axis-aligned rectangles, and \eqref{eq:z} represents zonotopes.

Throughout this paper, we refer to sets represented in the form of \eqref{eq:vrep} and \eqref{eq:hrep} as \emph{polytopes} and those in the form of \eqref{eq:cz} as constrained zonotopes, even though they all represent the same set~\cite{scott_constrained_2016,raghuraman2022set,vinod2024projection}.
We refer to unbounded sets of the form \eqref{eq:hrep} as \emph{polyhedra}.

Finally, we consider ellipsoidal sets $\cale$ \eqref{eq:ell}--\eqref{eq:ball}:
\begingroup
    \makeatletter\def\f@size{8.5}\check@mathfonts
\begin{align}
    \cale(Q,c) &= \{x\in\bbr^n\ |\ {(x-c)}^\top Q(x-c) \leq 1\}\label{eq:ell}\\
    \cale(G,c) &=\{x\in\bbr^n\ |\ \exists \xi\in\bbr^n,\ x=G\xi + c,\ \|\xi\|_2 \leq 1\}\label{eq:ell_gen}\\
    \calb(c,r) &= \{x\in\bbr^n\ |\ {\|x - c \|}_2\leq r\}\label{eq:ball}.
\end{align}
\endgroup
Here, the set representations \eqref{eq:ell} and \eqref{eq:ell_gen} are equivalent~\cite{boyd2004convex}, and \eqref{eq:ball} is a $n$-dimensional ball.

\subsection{Set operations}
For any sets $\calt, \cals\subseteq\bbr^n$ and $\calw\subseteq\bbr^m$, and a matrix $R\in\bbr^{m\times n}$, we define the set operations (affine map, Minkowski sum $\oplus$, intersection with inverse affine map $\cap_R$, and Pontryagin difference $\ominus$):
\begin{subequations}
\begin{align}
    R\calt &\triangleq\{R u: u \in \calc\},\label{eq:affinemap}\\
    \calt \oplus \cals &\triangleq\{u + v: u \in \calc,\ v\in\cals\},\label{eq:msum}\\
    \calt \cap_R \calw &\triangleq\{u\in\calc: Ru \in \calw\},\label{eq:intersection}\\
    \calt \ominus \cals &\triangleq\{u: \forall v \in\cals, u + v \in \calc\}.\label{eq:pdiff}
\end{align}\label{eq:set_operations}%
\end{subequations}
Since $\calc\cap\cals=\calc\cap_{I_n}\cals$, \eqref{eq:intersection} also includes the standard intersection. For any
$x\in\bbr^n$, we use $\calc + x$ and $\calc - x$ to denote $\calc \oplus \{ x\}$ and $\calc \oplus \{-x\}$ respectively
for brevity. 
The set operations \eqref{eq:set_operations} can also be used to define several other operations like orthogonal projection and inverse affine transformation (see Table~\ref{tab:methods}), and slicing (an intersection with an axis-aligned affine set).

The key advantage of constrained zonotopes over polytopes is that they admit closed form expressions for all set operations listed in \eqref{eq:set_operations} (except  Pontryagin difference \eqref{eq:pdiff})~\cite{scott_constrained_2016,raghuraman2022set}.
Recently, the authors have proposed a closed-form expression to inner-approximate the Pontryagin difference \eqref{eq:pdiff}~\cite{vinod2024projection}.
In contrast, polytopes must contend with computationally expensive vertex-halfspace enumeration when certain set operations are performed on a polytope in vertex/halfspace representation.

\section{The \pycvxset{} package}

\pycvxset{} provides three classes for representing convex sets \eqref{eq:vrep}--\eqref{eq:ball}: \polytope, \conzono, and \ellipsoid. 
In this section, we provide a brief overview of how we use \pycvxset{} to define, manipulate, and visualize these sets in Python.

\subsection{Set definitions}

\subsubsection{Polytope}
\label{sub:polytope_repr}
We define a set in polytopic representation using the \polytope{} class in the following ways:\\
1) specifying $(V)$ to define a polytope in V-Rep \eqref{eq:vrep},\\
2) specifying $(A,b)$ or $(A,b,Ae,be)$ to define a polytope in H-Rep \eqref{eq:hrep}, and\\
3) specifying rectangles $(l,u)$ (see \eqref{eq:rect}) or $(c,h)$ (see \eqref{eq:centered_rect}).
We also provide methods to convert a polytope from vertex representation to half-space representation and vice versa using \pycddlib{} and \scipy{}.

The following code snippet creates a polytope in V-Rep and $3$-dimensional simplex in H-Rep, prints the description of the polytope along with its vertices.

\begin{python}
import numpy as np
from pycvxset import Polytope

V = [[-1,0.5],[-1,1],[1,1],[1,-1],[0.5,-1]]
P1 = Polytope(V=V)
print("P1 is a", repr(P1))
A, b = -np.eye(3), np.zeros((3,))
Ae, be = [1, 1, 1], 1
P2 = Polytope(A=A, b=b, Ae=Ae, be=be)
print("P2 is a", P2)
print("Vertices of P2 are:\n", P2.V)
print("P2 is a", P2)
\end{python}%
The above code snippet produces the following output:
\begin{pythonoutput}
P1 is a Polytope in R^2 in only V-Rep
	In V-rep: 5 vertices
P2 is a Polytope in R^3 in only H-Rep
Vertices of P2 are:
 [[0. 0. 1.]
 [0. 1. 0.]
 [1. 0. 0.]]
P2 is a Polytope in R^3 in H-Rep and V-Rep
\end{pythonoutput}%
The call \texttt{P2.V} in Line 11 triggers a vertex enumeration internally as seen from the \texttt{print} statements for \texttt{P2}.

\subsubsection{Constrained zonotope}
\label{sub:cz_repr}
We define a set in constrained zonotopic representation \eqref{eq:cz} using the \conzono{} class in the following ways:\\
1) specifying $(c,G,Ae,be)$ as given in \eqref{eq:cz},\\
2) specifying $(c,G)$ as given in \eqref{eq:z} to define a zonotope,\\
3) specifying a \polytope{} object \eqref{eq:vrep}, \eqref{eq:hrep}, and\\
4) specifying rectangles $(l,u)$ (see \eqref{eq:rect}) or $(c,h)$ (see \eqref{eq:centered_rect}).

The following code snippet creates a constrained zonotope from the polytope defined before as well as a box.

\begin{python}
from pycvxset import ConstrainedZonotope

C1 = ConstrainedZonotope(polytope=P1)
print("C1 is a", repr(C1))
print("P1 is a", repr(P1))
C2 = ConstrainedZonotope(lb=[-1, -1], ub=[1, 1])
print("C2 is a", repr(C2))
\end{python}
The above code snippet produces the following output:
\begin{pythonoutput}
C1 is a Constrained Zonotope in R^2
	with latent dimension 7 and 5 equality constraints
P1 is a Polytope in R^2 in H-Rep and V-Rep
	In H-rep: 5 inequalities and no equality constraints
	In V-rep: 5 vertices
C2 is a Constrained Zonotope in R^2
	that is a zonotope with latent dimension 2
\end{pythonoutput}
\pycvxset{} performs a halfspace enumeration for \texttt{P1} (currently in vertex representation) in order to generate a constrained zonotopic representation \eqref{eq:cz} using~\cite[Thm. 1]{scott_constrained_2016}.
\pycvxset{} also detects that \texttt{C2} is a zonotope.

\pycvxset{} provide methods to generate polytopic approximations of constrained zonotopes (see Section~\ref{sub:plot}). Due to the computational effort involved~\cite{scott_constrained_2016}, an exact conversion from \eqref{eq:cz} to \eqref{eq:vrep} or \eqref{eq:hrep} is not supported.

\subsubsection{Ellipsoid}
We define an ellipsoid using the \ellipsoid{} class in the following ways:\\
1) specifying $(Q,c)$ as given in \eqref{eq:ell},\\
2) specifying $(G,c)$ as given in \eqref{eq:ell_gen}, and\\
3) specifying $(c,r)$ to define a ball \eqref{eq:ball}.\\

\pycvxset{} currently supports only ellipsoids with positive definite $Q$, i.e., full-dimensional ellipsoids (see~\cite{boyd2004convex}).
The following code snippet creates two ellipsoids of the forms \eqref{eq:ell} and \eqref{eq:ell_gen}.

\begin{python}
from pycvxset import Ellipsoid

E1 = Ellipsoid(c=[2,-1], Q=np.diag([1,4]))
print("E1 is an", E1)
E2 = Ellipsoid(c=[0,1,0], G=np.diag([1,2,3]))
print("E2 is an", E2)
\end{python}
The above code snippet produces the following output:
\begin{pythonoutput}
E1 is an Ellipsoid in R^2
E2 is an Ellipsoid in R^3
\end{pythonoutput}

\subsection{Visualizing polytopes and polytopic approximations}
\label{sub:plot}

\begin{figure}[!h]
    \centering
    \includegraphics[width=0.8\linewidth, trim={90 120 80 130}, clip]{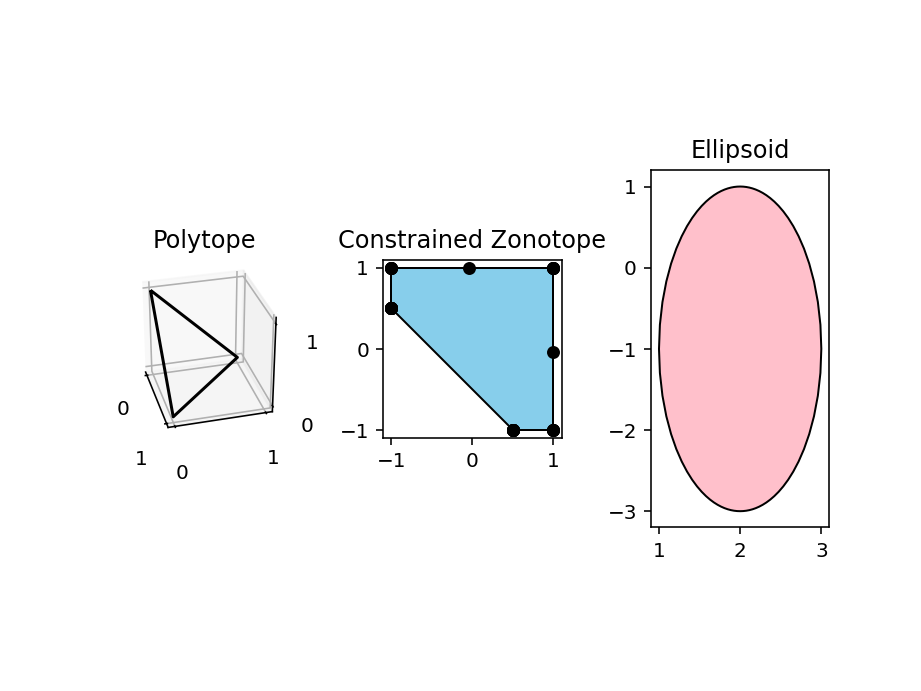}
    \caption{Plotting various sets using \pycvxset{}.}\label{fig:set_plots}
\end{figure}

We plot 2D and 3D polytopes using \pycvxset{} and \matplotlib, and plot polytopic approximations of constrained zonotopes and ellipsoids (see Fig.~\ref{fig:set_plots}). 

The following code snippet plots the sets in Fig.~\ref{fig:set_plots}.
\begin{python}
import matplotlib.pyplot as plt

plt.figure()
ax = plt.subplot(131, projection="3d")
P2.plot(ax=ax)  # Plot polytope
ax.view_init(elev=30, azim=-15)
ax.set_aspect("equal")
ax.set_title("Polytope")
ax = plt.subplot(132)  # Plot const. zonotope
C1.plot(ax=ax, vertex_args={"visible": True})
ax.set_aspect("equal")
ax.set_title("Constrained Zonotope")
ax = plt.subplot(133)  # Plot ellipsoid
E1.plot(ax=ax,patch_args={"facecolor":"pink"})
ax.set_aspect("equal")
ax.set_title("Ellipsoid")
plt.subplots_adjust(wspace=0.5)
\end{python}
\pycvxset{} provides flexibility in plotting either faces, vertices, or both, and provides identical methods for plotting irrespective of the set representation.
By default, \pycvxset{} plots inner-approximations of constrained zonotopes and ellipsoids, but outer-approximations may be plotted when required. 
For brevity, we will omit plotting commands in subsequent code snippets.

\begin{figure}
    \centering
    \includegraphics[width=0.4\linewidth, trim={190 30 130 80},clip]{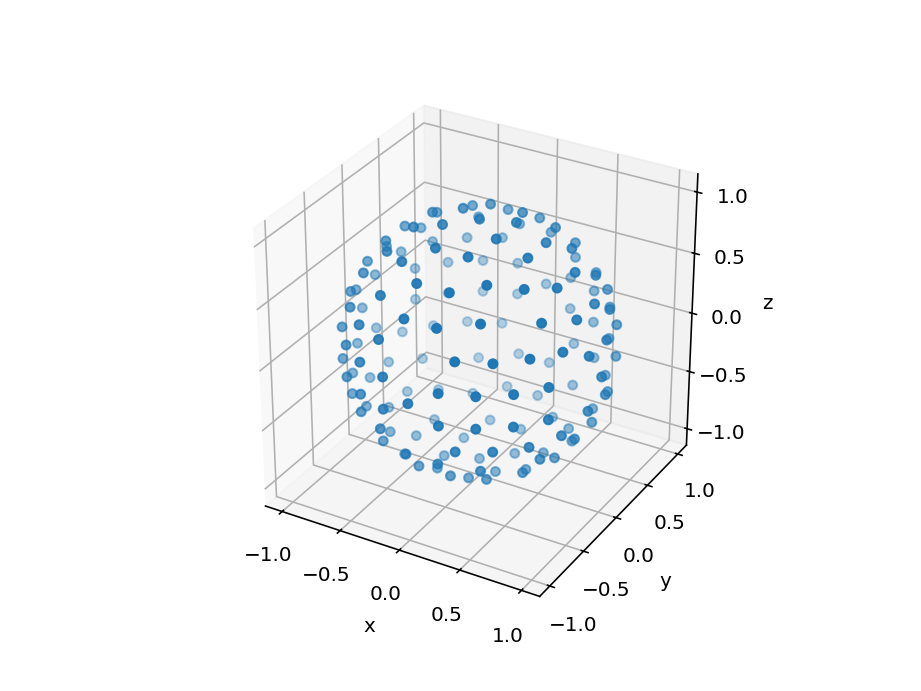}
    \caption{Well-separated vectors on a $3$-dimensional unit sphere.}\label{fig:ray_shoot}
\end{figure}

We compute polytopic inner- and outer-approximations for ellipsoids and constrained zonotopes of any $n$-dimensional set using their support function and support vectors. 
While \pycvxset{} accepts user-specified directions, it can also autogenerate well-separated $2n + 2^n D$ direction vectors for any $D\in\bbn$, by solving the following optimization problem~\cite[Eq. (B.1)]{gleason2021lagrangian},
\begingroup
    \makeatletter\def\f@size{8}\check@mathfonts
\begin{align}
    \hspace*{-1.5em}\begin{array}{rll}
        \text{max.} & r\\
        \text{s.\ t.} & {\|x_i - x_j\|}_2 \geq r,&\forall 1\leq i<j\leq D,\\
                           & {\|x_i - e_j\|}_2 \geq r,&\forall 1\leq i\leq D, \ \forall 1\leq j\leq n,\\
                           & 2x_i \geq r,\ 0.8 \leq \|x_i\| \leq 1,&\forall 1\leq i\leq D.\\
    \end{array}\label{eq:prob_spoaus}
\end{align}
\endgroup
Here, the decision variables are vectors $x_i\in\bbr^n$ for $i\in\{1,2,\ldots,D\}$ and a scalar $r$, and 
$e_j$ denotes the standard axis vector in $\bbr^n$.
\eqref{eq:prob_spoaus} is a difference-of-convex program that aims to spread points $x_i$ on the intersection of a unit sphere and the positive quadrant $\bbr_{\geq 0}^n$, which are subsequently reflected the axis planes to yield the direction vectors~\cite{gleason2021lagrangian,SReachTools}. \eqref{eq:prob_spoaus} may be solved (approximately) via the well-known \emph{convex-concave} procedure~\cite{lipp2016variations}, and the approach is implemented in \pycvxset{} as the method \texttt{spread\_points\_on\_a\_unit\_sphere}. 
\pycvxset{} uses $D=20$ as the default value.
Fig.~\ref{fig:ray_shoot} shows the result of \eqref{eq:prob_spoaus} for $n=3$.

\subsection{Set operations}
\begin{table*}[!th]
\centering
\caption{\pycvxset{} set operations on a set $X\subset\bbr^n$. Here, \cmark{} indicates exact implementation, $\changepolytopetocz$ indicates exact implementation possible after converting \polytope{} to \conzono{}, 
and $\approx$ indicates approximate implementation. All operations for constrained zonotopes and ellipsoids may be approximated, if desired, using appropriate polytopic approximations.}\label{tab:methods}
\adjustbox{width=1\linewidth}{\begin{tblr}{|l|c||c|c|c|}
   \hline
   \SetCell[r=2]{m} {Operation} & \SetCell[r=2]{m}{Expression} & \SetCell[r=2]{m}{Polytope} & Constrained & \SetCell[r=2]{m}{Ellipsoid} \\
    &  & & zonotope & \\\hline\hline
    \SetCell[c=5]{c} {Set computations involving another vector $v$ and/or matrix $M$} \\\hline\hline
    Affine transformation $(M,v)$  for $M\in\bbr^{m\times n},v\in\bbr^m$ & $MX + v$& \cmark & \cmark & \cmark\ ($M$ full row-rank)\\\hline
    Inverse-affine transformation $M\in\bbr^{n\times n}$, $M$ is invertible & $\{x\ |\ Mx \in X\}$ & \cmark & \cmark & \cmark\\\hline
    Project a point $v\in\bbr^n$ on to $X$  using ${\|\cdot\|}_p$, $p\in\{1,2,\infty\}$ & $\arg\min_{x\in X} \|x-v\|_p$ & \cmark & \cmark & \cmark \\\hline 
    Containment of a point $v\in\bbr^n$ in $X$ & $v\in X$ & \cmark & \cmark & \cmark \\\hline
    Support function along a direction $v\in\bbr^n$ & $\sup_{x\in X} v^Tx$ & \cmark & \cmark & \cmark \\\hline
    Support vector along a direction $v\in\bbr^n$ & $\arg\sup_{x\in X} v^Tx$ & \cmark & \cmark & \cmark \\\hline\hline
    \SetCell[c=5]{c} {Centering}\\\hline\hline
    Chebyshev center & $\sup_{\mathrm{Ball}(x, r)\subseteq X}\ r$ & \cmark & subopt. $\approx$  & \cmark \\\hline
    Maximum volume inscribed ellipsoid & $\sup_{\cale(c, Q)\subseteq X}\ \mathrm{Vol}(\cale)$ & \cmark &  subopt. $\approx$ & \cmark \\\hline
    Minimum volume circumscribed ellipsoid & $\inf_{\cale(c, Q)\supseteq X}\ \mathrm{Vol}(\cale)$ & \cmark & & \cmark \\\hline
    Minimum volume circumscribed rectangle & $\inf_{\mathrm{Rect}(l, u) \supseteq X}\ \mathrm{Vol}(\mathrm{Rect})$ & \cmark & \cmark & \cmark \\\hline\hline
    \SetCell[c=5]{c} {Other set-specific manipulations/computations}\\\hline\hline
    Interior point (Relative) & Compute $x\in X$ & \cmark & \cmark & \cmark \\\hline
    Orthogonal projection to $\bbr^m$ & $\{x | \exists v\in\bbr^{n-m},\ [x;v]\in X \}$ & \cmark & \cmark & \cmark \\\hline
    Volume & $\mathrm{Vol}(X)$ & \cmark & $\approx$ ($n=2$) & \cmark \\\hline\hline
    \SetCell[c=5]{c} {Set computations involving another set $Y$ ($Y\subset\bbr^n$ unless specified otherwise)} \\\hline\hline
    Intersection with a polytope $Y$ & \SetCell[r=4]{m} {$X \cap Y$} & \cmark & \cmark & \\\cline{1-1}\cline{3-5}
    Intersection with a constrained zonotope $Y$ & & $\changepolytopetocz$ & \cmark & \\\cline{1-1}\cline{3-5}
    Intersection with an affine set $Y$ & & \cmark & \cmark &  \\\cline{1-1}\cline{3-5}
    Intersection with a polyhedron $Y$ & & \cmark & \cmark & \\\hline
    Intersection with $Y\subset\bbr^m$ under inverse affine map $M\in\bbr^{m\times n}$ & {$X\cap_M Y=\{x\in X|Mx\in Y\}$} & {\cmark} & {\cmark} & \\\hline
    Minkowski sum with a polytope $Y$ & \SetCell[r=2]{m} {$X \oplus Y$} & \cmark & \cmark &  \\\cline{1-1}\cline{3-5}
    Minkowski sum with a constrained zonotope $Y$ & & $\changepolytopetocz$ & \cmark &  \\\cline{1-5}
    Pontryagin difference with an ellipsoid $Y$ & \SetCell[r=3]{m} {$X \ominus Y$} & \cmark & inner $\approx$  & \\\cline{1-1}\cline{3-5}
    Pontryagin difference with a zonotope $Y$ & & \cmark & inner $\approx$ & \\\cline{1-1}\cline{3-5}
    Pontryagin difference with a polytope $Y$ & & \cmark & & \\\hline
    Containment of a polytope $Y$  & \SetCell[r=3]{m} {$Y\subseteq X$} & \cmark & \cmark & \cmark \\\cline{1-1}\cline{3-5}
    Containment of a constrained zonotope $Y$ & & \cmark & \cmark & \\\cline{1-1}\cline{3-5}
    Containment of an ellipsoid $Y$  &  & \cmark &  & \cmark \\\hline
\end{tblr}}
\end{table*}

Table~\ref{tab:methods} lists the set operations supported for each class.
\pycvxset{} provides identical methods for various set operations (when supported).

\subsubsection{Involving another vector $v$ and/or matrix $M$}
For any set $X$, we support affine transformation $(M,v)$, inverse-affine transformation with invertible map $M$, projecting a point $v$, checking if $v\in X$, and computing the support function and vector of the set $X$ along the direction $v$.
We require inverse-affine map to have an invertible $M$ to ensure that the pre-image of a bounded set under the affine map $M$ is also bounded.
For an affine transformation of an ellipsoid, we require $M$ used to have full row-rank to guarantee full-dimensionality of the image ellipsoid~\cite{Elltoolbox}.
These operations either have a closed-form expressions (e.g., affine transformation of constrained zonotope~\cite{scott_constrained_2016} or support function of an ellipsoid~\cite{Elltoolbox}) or require solving convex programs which we implement using \cvxpy{} (e.g., checking $v\in X$ for a polytope $X$ is a linear program).

\begin{figure}[!h]
    \centering
    \includegraphics[width=0.8\linewidth, trim={70 170 270 250}, clip]{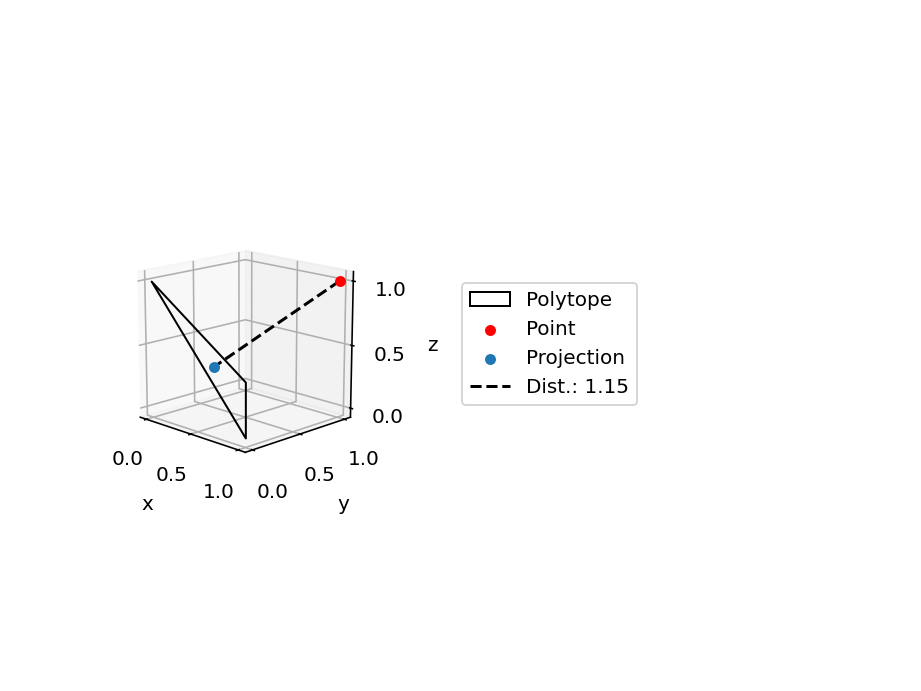}
    \caption{Projection of a point on a polytope}\label{fig:proj}
\end{figure}

As an illustration, we compute the Euclidean projection and the distance of a point $[1, 1, 1]$ on the polytope \texttt{P2} using the following code snippet (see Fig.~\ref{fig:proj}).
\begin{python}
point = np.array([1, 1, 1])
projection, d = P2.project(point, p=2)
\end{python}

\subsubsection{Centering} 
Centering methods provide a succinct, approximate representation of complex sets in the form of ellipsoids and rectangles~\cite[Ch. 8]{boyd2004convex}.
These methods solve convex programs for polytopes in V-Rep/H-Rep, and are available in closed-form for ellipsoids~\cite{Elltoolbox}.
For constrained zonotopes, we provide approximations~\cite{vinodCDC2024}.

\begin{figure}[!h]
    \centering
    \includegraphics[width=1\linewidth, trim={90 210 90 220}, clip]{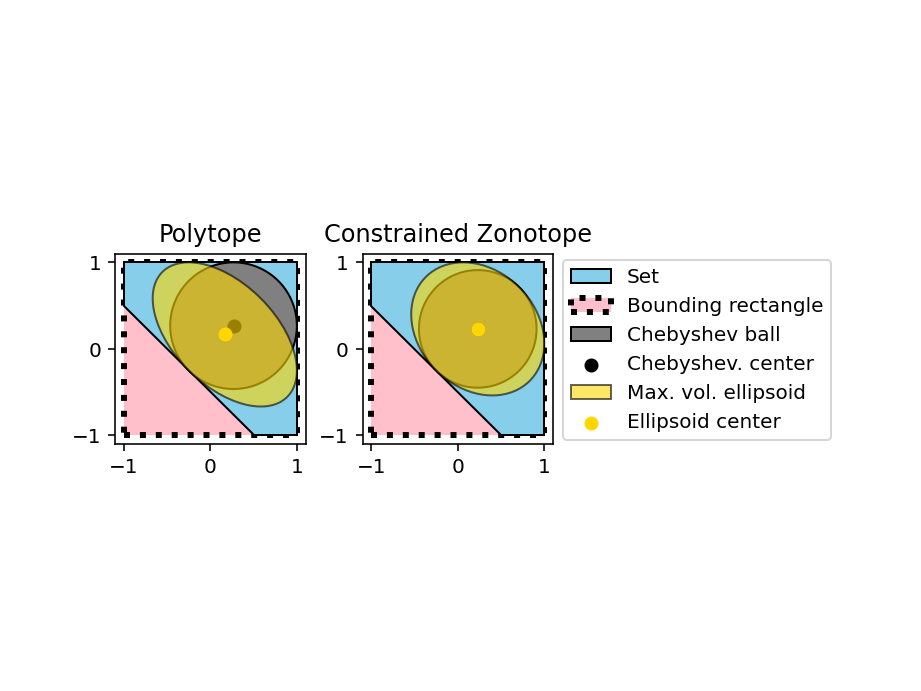}
    \caption{Centering for a polytope and a constrained zonotope.}\label{fig:centering}
\end{figure}

Fig.~\ref{fig:centering} illustrates centering and bounding sets for the polytope \texttt{P1} and the constrained zonotope \texttt{C1}.
We obtained Chebyshev radius and ellipsoids of volume of $0.73$ and $1.89$ for \texttt{P1} and $0.68$ and $1.81$ for \texttt{C1} respectively, and obtained identical rectangles.

\subsubsection{Other set-specific manipulations/computations}

We use centering techniques for the computation of an interior point for each set. 
For polytopes, we can also compute its centroid (the mean of its vertices) as an alternative.

We compute the orthogonal projection of a set using an appropriately defined affine map.

\begin{figure}[!h]
    \centering
    \includegraphics[width=0.8\linewidth, trim={95 120 90 160}, clip]{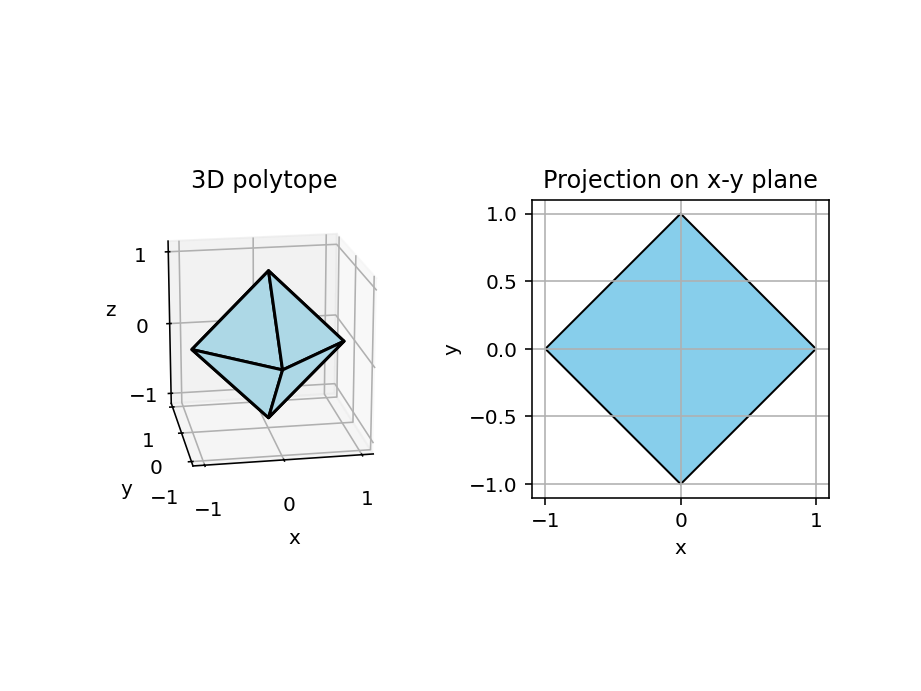}
    \caption{Projection of a $3$-dimensional unit $\ell_1$-norm ball.}\label{fig:set_projection}
\end{figure}

We compute the projection of a $3$-dimensional unit $\ell_1$-norm ball using the following code snippet (see Fig.~\ref{fig:set_projection}). 
Note that \pycvxset{} counts dimensions from zero.
\begin{python}
V = np.vstack((np.eye(3), -np.eye(3)))
l1ball = Polytope(V=V)
ball2D = l1ball.projection(project_away_dim=2)
\end{python}

We compute the volume of a full-dimensional polytope and an ellipsoid using \texttt{scipy} and closed-form expressions respectively.
We approximate the volume of a constrained zonotope via grid-based sampling.

\subsubsection{Involving another set $Y$}
\label{sub:involving_set}

We provide exact implementations for intersection and Minkowski sum of polytopes and constrained zonotopes among themselves, and for Pontryagin difference of polytopes with any other sets~\cite{gilbert}. 
The intersection and Minkowski sum of a constrained zonotope and a polytope returns a constrained zonotope.  
The Pontryagin difference of a constrained zonotope and an ellipsoid or a zonotope is inner-approximated with a constrained zonotope using least squares~\cite{vinod2024projection}.

We implement an exact check for the containment of a polytope $Y$ within a set $X$ (not necessarily a polytope) by solving appropriate convex programs~\cite{boyd2004convex}, where we check for the containment of all the vertices of $Y$ in the set $X$.
We also implement an exact check for the containment of a set $Y$ (not necessarily a polytope) within a polytope $X$ using the support function~\cite{boyd2004convex}.
We implement an exact check for the containment of an ellipsoid in another ellipsoid using semi-definite programming~\cite{boyd2004convex}.

To implement exact check for $Y\subseteq X$ for two constrained zonotopes $X,Y$, we encode $x\in Y\Rightarrow x\in X$ as a bilinear program obtained using strong duality:
\begingroup
    \makeatletter\def\f@size{9}\check@mathfonts
\begin{align}
    \begin{array}{rl}
        \text{minimize} & 1 + \alpha^\top(c_Y - G_X\xi_X - c_X) - \beta^\top b_{e, Y}\\
        \text{subject\ to} & \|\xi_X\|_\infty \leq 1,\\
    & A_{e,X} \xi =b_{e,X},\\
    &{\|G_Y^\top \alpha + A_{e,Y}^\top \beta\|}_1 \leq 1.
    \end{array}\label{eq:containment_bilinear}
\end{align}
\endgroup
with decision variables $\alpha\in\bbr^n,\beta\in\bbr^{M_Y},\xi_X\in\bbr^{N_X}$, and $Y\subseteq X$ if and only if the optimal value of \eqref{eq:containment_bilinear} is non-negative.
We solve \eqref{eq:containment_bilinear} to optimality using \cvxpy{} and \gurobi{}, and can also check for containment of polytopes within constrained zonotopes.
These methods also enable checking for equality between polytopes and constrained zonotopes, as illustrated in the following code snippet.

\begin{python}
print('C1 is a', C1)
print('P1 is a', P1)
print('Are C1 and P1 equal?', C1 == P1)
lb, ub, p, q = [-1, -1], [1, 1], [-1,-1], 0.5
P1a = Polytope(lb=lb, ub=ub).intersection_with_halfspaces(A=p, b=q)
C1a = ConstrainedZonotope(lb=lb, ub=ub).intersection_with_halfspaces(A=p, b=q)
print('C1a is a', C1a)
print('P1a is a', P1a)
print('Are P1 and P1a equal?', P1a == P1a)
print('Are C1 and C1a equal?', C1 == C1a)
\end{python}
The above code snippet produces the following output:
\begin{pythonoutput}
C1 is a Constrained Zonotope in R^2
P1 is a Polytope in R^2 in H-Rep and V-Rep
Are C1 and P1 equal? True
C1a is a Constrained Zonotope in R^2
P1a is a Polytope in R^2 in only H-Rep
Are P1 and P1a equal? True
Are C1 and C1a equal? True
\end{pythonoutput}
The equality of the sets \texttt{C1} and \texttt{P1} may also be  visually confirmed in Fig.~\ref{fig:centering}, where the polytopic inner-approximation of \texttt{C1} computed by \pycvxset{} for plotting is exact for this constrained zonotope instance.
In contrast to the sets \texttt{P1} and \texttt{C1} defined in Sections~\ref{sub:polytope_repr} and~\ref{sub:cz_repr}, the sets \texttt{P1a} and \texttt{C1a} defined in Lines 5 and 6 in the above code snippet are defined by an intersection of a unit $\ell_\infty$-norm ball and an appropriate halfspace $\{x:p^\top x \leq q\}$.
As expected, \pycvxset{} declares all these sets to be equal, despite being different representations.

We support intersection of \polytope{} and \conzono{} objects with unbounded sets like affine sets and polyhedron since these operations are also closed in \polytope{} and \conzono{} respectively. We also implement \texttt{slice} using intersection with an appropriately-defined affine set.

We do not support intersection, Minkowski sum, and Pontryagin difference operations for ellipsoids natively in \pycvxset{} since they are not closed operations for ellipsoids. 
However, all set operations discussed here are supported by \polytope{}.
Consequently, any set operation that is not natively supported by \pycvxset{} involving constrained zonotopes and ellipsoids may be approximated using their appropriate polytopic approximations (Section~\ref{sub:plot}).
In Table~\ref{tab:methods}, \cmark{} indicates implementations of set operations in \pycvxset{} that yields an object of the same class as $X$ and does not rely on polytopic approximations.

\subsection{Overloaded operators}

We overload several Python operators to simplify the use of \pycvxset{}. Table~\ref{tab:operators} summarizes how these operators interact with the sets in \pycvxset{}. 

When the comparison operators ({\codestyle <,<=,>,>=,\textbf{in}}) are given a vector $y\in\bbr^n$ instead of a set $Y\subset\bbr^n$, \pycvxset{} automatically switches to appropriate containment check with the vector $y$. 
Similarly, when the addition/subtraction operator is given a vector $y\in\bbr^n$, $X+y$, $y+X$, and $X-y$ translates $X$ by $y$, $y$, and $-y$ respectively.

We also support the operation of Cartesian product of a \polytope{} or a \conzono{} with itself.
\begin{table}[!th]
\centering
\caption{Examples of python expressions involving sets $X,Y\subset\bbr^n$ defined from \pycvxset{}, and an appropriate matrix $M$}\label{tab:operators}
\adjustbox{width=1\linewidth}{
\begin{tblr}{|c|c|}
   \hline
Python expression & Interpretation \\\hline\hline
{\codestylenormal M * X, M @ X} & Affine map with $M\in\bbr^{m\times n}$\\\hline
{\codestylenormal -X} & Negation, equivalent to {\codestylenormal -np.eye(X.dim) @ M}\\\hline
{\codestylenormal X * M, X @ M} & Inverse affine map with invertible $M\in\bbr^{n\times n}$\\\hline
{\codestylenormal X < Y, X <= Y, X \textbf{in} Y} & $X\subseteq Y$ \\\hline
{\codestylenormal X > Y, X >= Y, Y \textbf{in} X} & $X\supseteq Y$ \\\hline
{\codestylenormal X == Y} & Equality check --- $X\subseteq Y$ and $Y\subseteq X$\\\hline
{\codestylenormal X + Y} & Minkowski sum of $X$ with set $Y$ \\\hline
{\codestylenormal X - Y} & Pontryagin difference of $X$  with set $Y$\\\hline
{\codestylenormal X ** m} & Cartesian product with itself $m\in\bbn$ times\\\hline
\end{tblr}}
\end{table}

\subsection{Solving relevant optimization problems}
We use \cvxpy{} to solve various optimization problems within \pycvxset{}.
We also provide methods to set up and solve convex programs with \cvxpy{} involving sets constructed using \pycvxset{}:\\
1) {\texttt{minimize}} to set up and solve optimization problems,
\begin{align}
    \begin{array}{rl} 
        \text{minimize} & J(x),\\
        \text{subject\ to} & x\in X,\\
    \end{array}\label{eq:opt}
\end{align}
for any \cvxpy{}-compatible cost function $J$, and\\
2) {\texttt{containment\_constraints}} to obtain the \cvxpy{} expressions that enforce the containment constraints $x\in X$ as well as any necessary auxiliary variables.\\
Various methods in \pycvxset{} like \texttt{project}, \texttt{support} use these methods to solve convex programs.

The user can specify the solver to use during set computations via the attributes \texttt{cvxpy\_args\_lp}, \texttt{cvxpy\_args\_socp}, and  \texttt{cvxpy\_args\_sdp} associated with each object.
These attributes are used when solving the various linear programs, second-order cone programs, and semi-definite programs respectively.

\subsection{Installation and examples}

The source code of \pycvxset{} along with detailed documentation and OS-specific installation instruction are available at \underline{\href{https://github.com/merlresearch/pycvxset}{https://github.com/merlresearch/pycvxset}}. 
We have tested \pycvxset{} in Windows, Ubuntu, and MacOS, and for Python versions from 3.9 to 3.12. In future, we plan to register \pycvxset{} to the Python Package Index as well. 
\pycvxset{} is released under the AGPL-3.0-or-later license.

\begin{figure}
    \includegraphics[width=1\linewidth, trim={0 10 0 0}, clip]{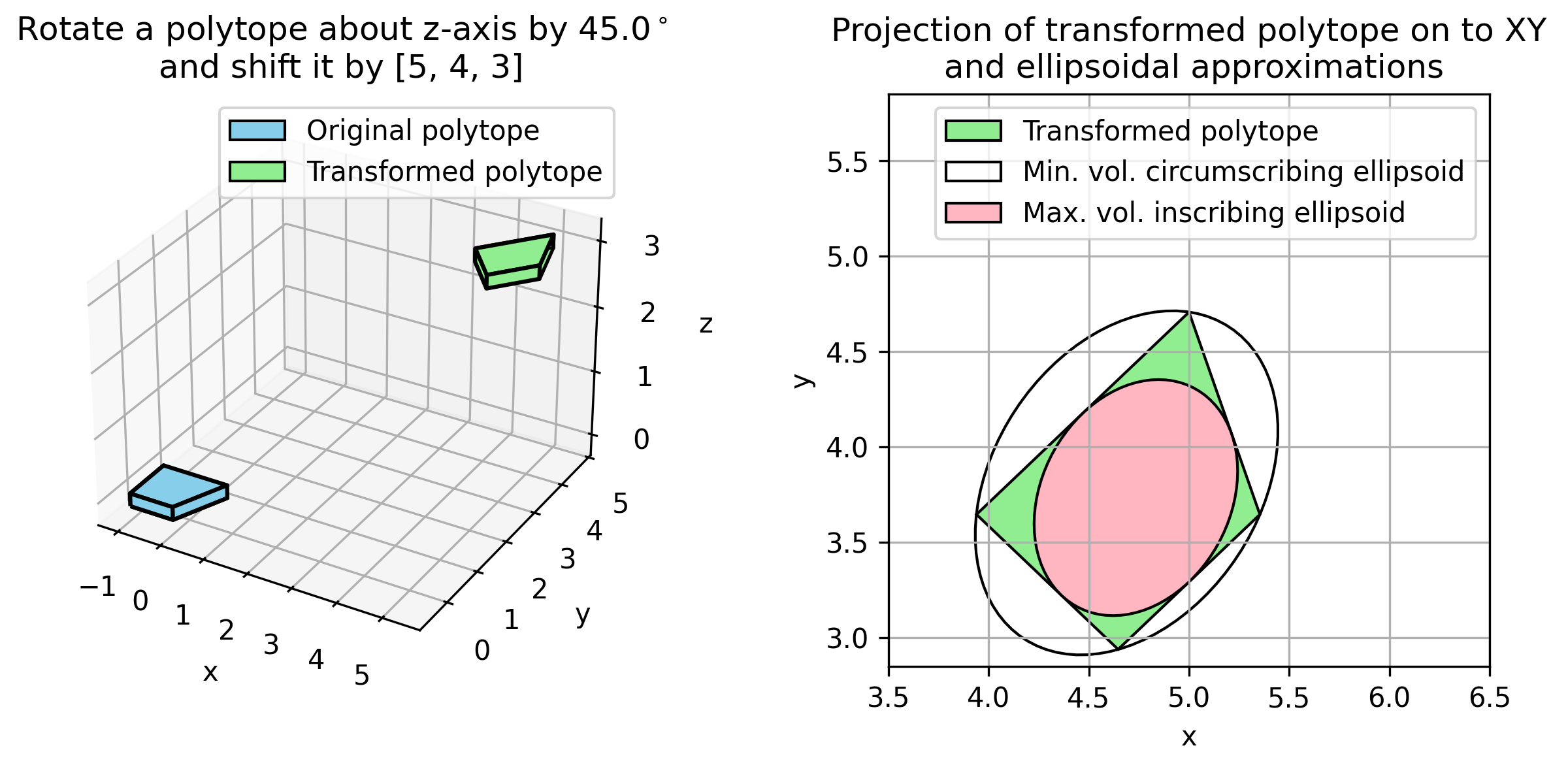}
    \caption{Figure generated by the script \texttt{pycvxset\_diag.py}.}\label{fig:pycvxset_diag}
\end{figure}

Additionally, we provide several Jupyter notebooks in the folder \texttt{examples/} to help a user understand the different functionalities of \pycvxset{}. We also provide a diagnostic script \texttt{pycvxset\_diag.py}, which can help in checking if \pycvxset{} was installed properly. Fig.~\ref{fig:pycvxset_diag} shows the results of running the command\vspace*{0.5em}\newline\indent\texttt{\$ python examples/pycvxset\_diag.py}\vspace*{0.5em}\newline from the root of the package.

\section{Reachability analysis using \pycvxset{}}
\label{sec:backward_recurse}

We now briefly discuss how \pycvxset{} may be used to compute robust controllable (RC) sets~\cite[Defn. 10.18]{borrelli2017predictive}.
Consider a discrete-time linear time-invariant system with additive uncertainty,
\begin{align}
    x_{t+1} = A x_t + B u_t + F w_t,\label{eq:dyn}
\end{align}
with state $x_t\in\bbr^n$, input $u_t\in\calu\subset\bbr^m$, disturbance $w_t\in\calw\subset\bbr^p$, and appropriate matrices $A,B,F$.
We assume that the input set $\calu$ and disturbance set $\calw$ are convex and compact sets.
Given a horizon $N\in\bbn$, a polytopic safe set $\cals\subset\bbr^n$ and a polytopic target set $\calt\subset\bbr^n$, a \emph{$N$-step robust controllable set} is the set of initial states that can be robustly driven, through a time-varying control law, to the target set in $N$ steps, while satisfying input and state constraints for all possible disturbances.
Formally, we define the $N$-step RC set as $\calk_0$ via the following set recursion for $t\in\{0, 1, \ldots, N-1\}$:
\begin{align}
    \calk_t = \cals \cap \left({A^{-1}{((\calk_{t+1} \ominus F\calw) \oplus (-B \calu))}}\right),\label{eq:backward_set_recursion}
\end{align}
with $\calk_N\triangleq \calt$. We implement \eqref{eq:backward_set_recursion} in \pycvxset{} with the following Python function \texttt{get\_rcs}.
\begin{python}
def get_rcs(S_U, S_W, S_S, S_T, A, B, F, N):
    S_K = [None] * (N + 1)
    S_K[-1], S_FW, S_BU = S_T, F@S_W, (-B)@S_U
    for t in range(N - 1, -1, -1):
        S_temp = (S_K[t+1] - S_FW) + S_BU
        S_K[t]=S_S.intersection(S_temp @ A)
    return S_K[0]
\end{python}
In \texttt{get\_rcs}, we highlight variables denoting sets with a prefix \texttt{S\_} to distinguish from other variables --- the horizon $N$ and matrices $A,B,F$ defined in \eqref{eq:dyn}.
Here, \texttt{S\_U} is $\calu$, \texttt{S\_W} is $\calw$, \texttt{S\_T} is $\calt$, and \texttt{S\_S} is $\cals$. 

Lines 2 and 3 of \texttt{get\_rcs} initialize the sets and pre-compute the affine-mapped sets in \eqref{eq:backward_set_recursion}.
Lines 5-6 implement the set recursion \eqref{eq:backward_set_recursion} using Table~\ref{tab:operators}. 
The returned set \texttt{S\_K[0]} is a \conzono{} (or a \polytope{}) when sets \texttt{S\_U}, \texttt{S\_T}, and \texttt{S\_S} are \conzono{} (or \polytope{}) objects. For a \conzono{}-based computation, the set \texttt{S\_W} must be a zonotope or an ellipsoid~\cite{vinod2024projection}, while the set \texttt{S\_W} can be any set in \pycvxset{} for the \polytope{}-based computation (see Table~\ref{tab:methods}).

\section{Numerical examples}

We provide several numerical examples to demonstrate various features of \pycvxset{}.  All computations were done on a
standard laptop with 13th Gen Intel i7-1370P, 20 cores, 64 GB RAM using Python 3.9.

\subsection{Reachability analysis for a double integrator}

We use \pycvxset{} to compute a $30$-step RC set for a double integrator system model.
A double integrator system can model an acceleration-controlled, mobile robot constrained to travel on a line.
The corresponding RC set indicates the safe initial states that allow for subsequent satisfaction of state and input constraints.
Using a sampling time of $0.1$, we have \eqref{eq:dyn} with,
\begin{align}
    A = \left[\begin{array}{cc}
            1 & 0.1\\
            0 & 1\\
        \end{array}
        \right], \ B = F = \left[\begin{array}{cc}
            0.005\\
            0.1\\
        \end{array}
        \right],\label{eq:double_intg}
\end{align}
with two-dimensional state $x_t$ denoting the position and velocity 
with one-dimensional input $u_t\in\mathcal{U}=[-1, 1]$ and disturbance $w_t\in\mathcal{W}=0.4\times\mathcal{U}$ denoting the controlled acceleration and the perturbation. 
We choose the safe set $\cals=[-1, 1]\times[-0.5, 0.5]$, which serves as position and velocity bounds the robot must satisfy at all times.
We choose the target set $\calt=[-0.25, 0.25]\times[-0.1, 0.1]$, which requires the robot to have a terminal position (at time $t=30$) within $0.25$ m of the origin, and a terminal velocity magnitude of at most $0.1$ m/s.

\begin{figure}[!h]
    \includegraphics[width=1\linewidth,trim={0 230 0 250},clip]{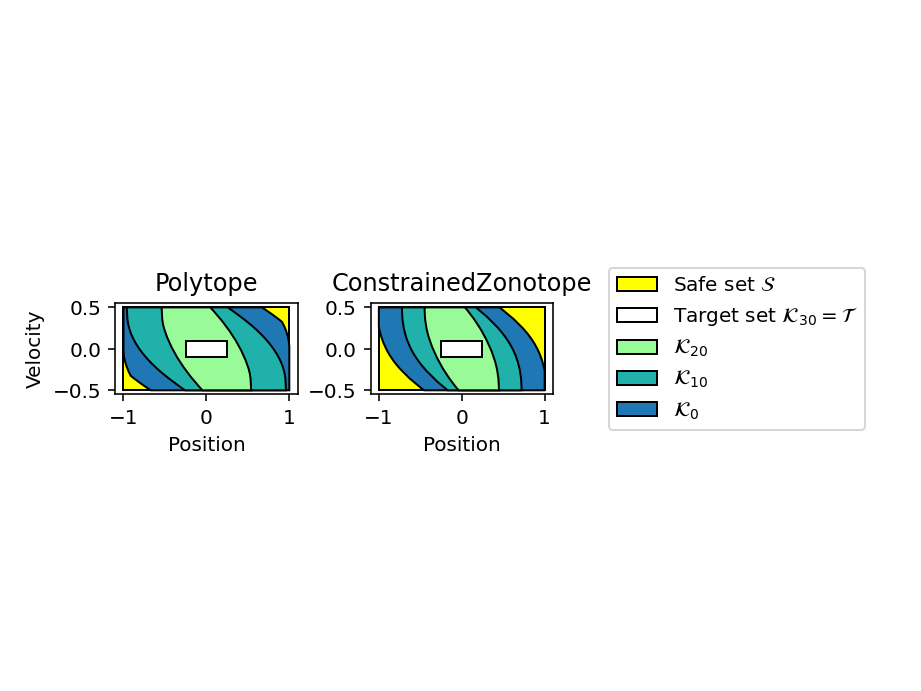}
    \caption{$30$-step RC sets for \eqref{eq:double_intg} using \pycvxset{}.}\label{fig:double_integrator}
\end{figure}

Fig.~\ref{fig:double_integrator} shows the RC sets computed using \texttt{get\_rcs}. 
Observe that the RC set computed using constrained zonotope is slightly smaller than the set computed using polytopes, due to the inner-approximation used in Pontryagin difference~\cite{vinod2024projection}.
The overall computation time to generate and plot Fig.~\ref{fig:double_integrator} was less than $3$ seconds.

\subsection{Reach-avoid computation for spacecraft rendezvous}

We now demonstrate a practical application of \pycvxset{} where \conzono{} class provides scalability and numerical stability over \polytope{} class for the computation of RC set.
It also uses an ellipsoidal uncertainty set defined using \ellipsoid{} class.

We consider the problem of safe spacecraft rendezvous. For safety, it is essential to characterize the set of safe terminal configurations from which an approaching spacecraft (deputy) may wait for go/no-go for docking with another spacecraft (chief)~\cite{gleason2021lagrangian,vinodCDC2024}.
From each of these positions, the deputy must be able to proceed towards the chief for docking using bounded control authority while staying within a line-of-sight cone and satisfying velocity bounds at all times.

\emph{Dynamics:} Assuming a circular orbit for the chief near the earth, the relative dynamics may be described by a four-dimensional linear system model, known as Hill-Clohessy-Wiltshire dynamics) 
to describe the position and velocity in relative \texttt{x}-\texttt{y} coordinates. 
We discretize the model in time using zero-order hold to obtain \eqref{eq:dyn} with $F$ set to a $4$-dimensional identity matrix~\cite{gleason2021lagrangian,vinodCDC2024,SReachTools}.
We assume that the thruster inputs $u_t\in\calu={[-0.2, 0.2]}^2$ N are held constant over the sampling time $30$ seconds.
We account for uncertainty in the rendezvous trajectory arising from potential actuator limitations of the spacecraft and model mismatch using an additive uncertainty $w_t\in\calw_t=\ellipsoid(c=[0, 0, 0, 0], G=[10^{-5}, 10^{-5}, 10^{-4}, 10^{-4}])\subset\bbr^4$ in the form \eqref{eq:ell_gen} (units km and m/s).

\emph{Computation of RC set:} We compute a $50$-step RC set to navigate the deputy to a target set $\calt={[-0.2, 0.2]}\times{[-0.2, 0]}\times{[-0.1, 0.1]}^2$ (units km and m/s).
Additionally, the deputy must remain inside a line-of-sight cone originating from the chief, 
$\cals=\{x\in\bbr^4: |x_1|\leq - x_2 \leq 1, |x_3|\leq 0.05, |x_4|\leq 0.05\}$ (units km and m/s). 
See~\cite{gleason2021lagrangian,vinodCDC2024} for more details.

\begin{figure}
    \includegraphics[width=1\linewidth,trim={0 20 0 100}, clip]{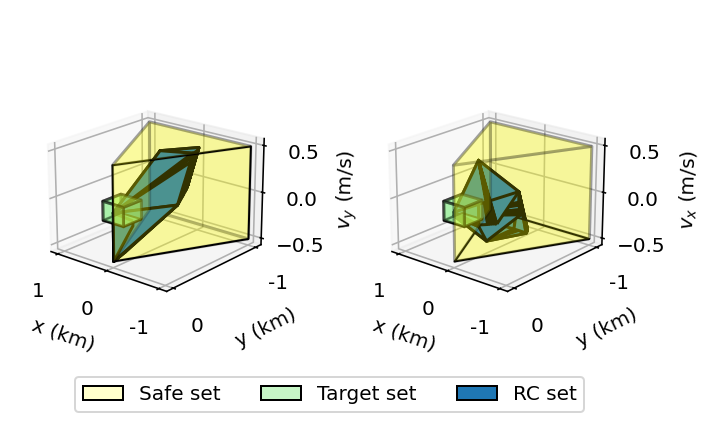}
    \caption{Slices of the $4$-dimensional $50$-step RC set for a spacecraft rendezvous problem. (Left) Initial $v_x=0$. (Right) Initial $v_y=0$.}\label{fig:cwh3D}
\end{figure}
Fig.~\ref{fig:cwh3D} shows the slices of the RC set computed using the \conzono{} class of \pycvxset{}.
We faced numerical issues when performing polytope-based computations of RC sets which may be attributed to the difficulties arising vertex-halfspace enumeration.
The computation of the RC set using \conzono{} took about $30$ seconds.

\subsection{Admissible deviations from a given robot trajectory}

We now demonstrate how \pycvxset{} and \conzono{} class may be used to compute projections and slices of high-dimensional polytopes.

Consider a robot moving in 2D space with (unperturbed) dynamics along each dimension given by a double integrator system \eqref{eq:double_intg}. While one can use convex optimization to design minimum-energy, dynamically-feasible, trajectories passing through a collection of waypoints, we consider the problem of characterizing all positions that the robot may \emph{deviate to} while following the trajectory using \pycvxset{}. 

For a horizon $N$, position waypoints $z_{i,\text{pos}}\in\bbr^2$ with $i\in\cali\subset\bbn$, dynamics $(A, B)$, initial state $x_0$, and input constraint set $\calu\subset\bbr^2$, we define the set of admissible trajectories $\cald\subset\bbr^{4N}$ as follows,
\begingroup
    \makeatletter\def\f@size{8}\check@mathfonts
\begin{align}
    \hspace*{-1em}\cald = \left\{
        \left[\begin{array}{c}
        x_1\\ \vdots\\ x_N\end{array}
        \right]\in\bbr^{4N} \middle| 
    \begin{array}{c}
    \forall t\in\{0,1,\ldots,N-1\},\\
    \exists u_t\in\calu,\ x_{t+1} = A x_t + B u_t,\\
    \forall i\in\cali,\ [I_2, 0_{2\times 2}]x_i = z_{i,\text{pos}}\\
    \end{array}\right\}.
\end{align}
\endgroup
We compute $\cald$ in \pycvxset{} via an affine transformation of the open-loop control sequence set $\calu^N$, where the transformation $(M,v)$ (see Table~\ref{tab:methods}) is characterized by $x_0, A, B, N$. 
We then slice the transformed set at appropriate dimensions to enforce the position of robot at time $i$ is $z_{i,\text{pos}}$ for each $i\in\cali$.
In other words, the set $\cald$ is a slice of a \emph{forward reachable set}~\cite{borrelli2017predictive}.
While such computations are numerically challenging for polytopes due to the  vertex-halfspace enumeration involved for high-dimensional polytopes, these sets are easily characterized using constrained zonotopes.

\begin{figure}
    \includegraphics[width=0.95\linewidth,trim={10 40 0 110}, clip]{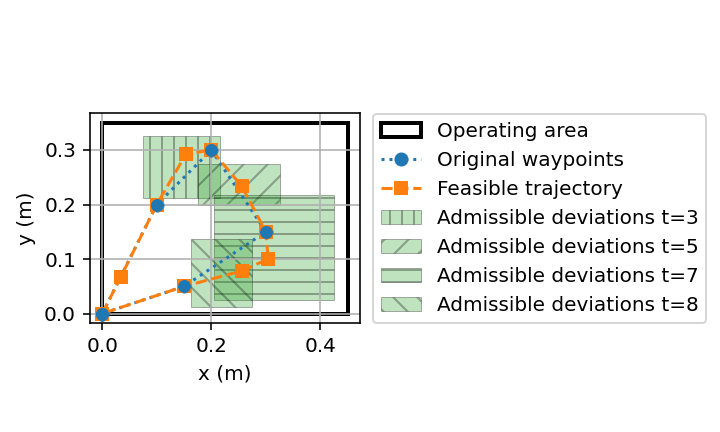}
    \includegraphics[width=0.95\linewidth,trim={10 40 0 110}, clip]{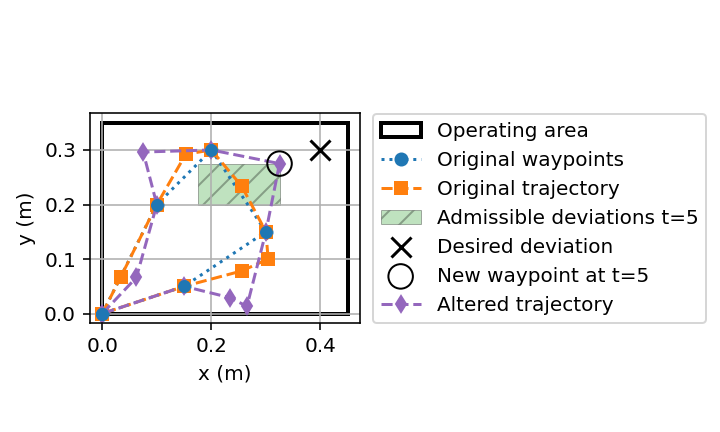}
    \caption{(Top) Original waypoints, feasible robot trajectory passing through the waypoints, and the admissible deviations at different time steps. While the first few and the penultimate time steps do not admit any deviations, the rest of the time steps (not in $\cali$) admit deviations. (Bottom) We generate a new waypoint from the admissible deviation set at time step $t=5$, and alter the robot trajectory to pass through the new waypoint.}\label{fig:deviation}
\end{figure}
Fig.~\ref{fig:deviation} (top) illustrates the set of admissible deviation positions and a dynamically feasible trajectory passing through these waypoints for $N=10$, dynamics with sampling time as $0.5$ seconds, and a set of waypoint position constraints at $\cali=\{2, 4, 6, 9\}$. 
We choose initial state $x_0=[0, 0, 0, 0]$, and input constraint set $\calu=\{u\in\bbr^2: -1\leq u \leq 1\}$ (in ms$^{-2}$). 
For ease in plotting, we defined separate sets of admissible deviation positions for each time $t\in\{1,3,5,7,8\}$, and restricted the velocity at these deviation positions to be zero. 
For the given set of waypoints, we found that no such admissible waypoint exists for $t=1$.
The set of all such admissible deviation positions over any time step covers an area of about $0.067$ m$^2$, which is around $42\%$ of the operating area $[0, 0.45]\times[0, 0.35]$ (in m).

Fig.~\ref{fig:deviation} (bottom) illustrates an altered trajectory to reach close to $(0.4, 0.3)$ at $t=5$. 

\section{Conclusion}
This paper introduces \pycvxset{}, an open-source Python package to manipulate and visualize convex sets in Python.
Currently, \pycvxset{} supports polytopic, ellipsoidal, and constrained zonotopic set representations.
The packages facilitates the use of set-based methods to analyze and control dynamical systems in Python.

\section{Acknowledgements}
We are grateful to Stefano Di Cairano and Kieran Parsons for their insightful feedback during the course of the development of this package.

\bibliographystyle{IEEEtran}
\bibliography{refs}

\begin{thebibliography}{10}
\providecommand{\url}[1]{#1}
\csname url@samestyle\endcsname
\providecommand{\newblock}{\relax}
\providecommand{\bibinfo}[2]{#2}
\providecommand{\BIBentrySTDinterwordspacing}{\spaceskip=0pt\relax}
\providecommand{\BIBentryALTinterwordstretchfactor}{4}
\providecommand{\BIBentryALTinterwordspacing}{\spaceskip=\fontdimen2\font plus
\BIBentryALTinterwordstretchfactor\fontdimen3\font minus
  \fontdimen4\font\relax}
\providecommand{\BIBforeignlanguage}[2]{{%
\expandafter\ifx\csname l@#1\endcsname\relax
\typeout{** WARNING: IEEEtran.bst: No hyphenation pattern has been}%
\typeout{** loaded for the language `#1'. Using the pattern for}%
\typeout{** the default language instead.}%
\else
\language=\csname l@#1\endcsname
\fi
#2}}
\providecommand{\BIBdecl}{\relax}
\BIBdecl

\bibitem{blanchini2008set}
F.~Blanchini and S.~Miani, \emph{Set-theoretic methods in control}.\hskip 1em
  plus 0.5em minus 0.4em\relax Springer, 2008.

\bibitem{bertsekas_1971}
D.~Bertsekas and I.~Rhodes, ``On the minimax reachability of target sets and
  target tubes,'' \emph{Automatica}, vol.~7, 1971.

\bibitem{borrelli2017predictive}
F.~Borrelli, A.~Bemporad, and M.~Morari, \emph{Predictive control for linear
  and hybrid systems}.\hskip 1em plus 0.5em minus 0.4em\relax Cambridge Univ.
  Press, 2017.

\bibitem{althoff2021set}
M.~Althoff, G.~Frehse, and A.~Girard, ``Set propagation techniques for
  reachability analysis,'' \emph{Annual Rev. Ctrl., Rob., \& Auto. Syst.},
  vol.~4, no.~1, pp. 369--395, 2021.

\bibitem{vinod2021abort}
A.~Vinod, A.~Weiss, and S.~Di~Cairano, ``Abort-safe spacecraft rendezvous under
  stochastic actuation and navigation uncertainty,'' in \emph{Proc. Conf. Dec.
  \& Ctrl.}, 2021, pp. 6620--6625.

\bibitem{vinod2024projection}
------, ``Projection-free computation of robust controllable sets with
  constrained zonotopes,'' \emph{arXiv preprint arXiv:2403.13730}, 2024.

\bibitem{vinodCDC2024}
------, ``Inscribing and separating an ellipsoid and a constrained zonotope:
  Applications in stochastic control and centering,'' in \emph{Proc. Conf. Dec.
  \& Ctrl.}, 2024, (accepted).

\bibitem{scott_constrained_2016}
J.~Scott, D.~Raimondo, G.~Marseglia, and R.~Braatz, ``Constrained zonotopes:
  {A} new tool for set-based estimation and fault detection,''
  \emph{Automatica}, vol.~69, pp. 126--136, 2016.

\bibitem{raghuraman2022set}
V.~Raghuraman and J.~Koeln, ``Set operations and order reductions for
  constrained zonotopes,'' \emph{Automatica}, 2022.

\bibitem{yang2022efficient}
L.~Yang, H.~Zhang, J.~Jeannin, and N.~Ozay, ``Efficient backward reachability
  using the {M}inkowski difference of constrained zonotopes,'' \emph{IEEE Tran.
  Comp.-Aided Design Integrated. Circ. \& Syst.}, vol.~41, no.~11, pp.
  3969--3980, 2022.

\bibitem{Elltoolbox}
A.~Kurzhanskiy and P.~Varaiya, ``Ellipsoidal toolbox {(ET)},'' in \emph{Proc.
  Conf. Dec. \& Ctrl.}, 2006,
  \url{http://systemanalysisdpt-cmc-msu.github.io/ellipsoids/}.

\bibitem{pycddlib}
``pycddlib,'' \url{https://pypi.org/project/pycddlib/}.

\bibitem{pytope}
``pytope,'' \url{https://github.com/heirung/pytope}.

\bibitem{polytope}
``polytope,'' \url{https://tulip-control.github.io/polytope/}.

\bibitem{pypoman}
``pypoman,'' \url{https://pypi.org/project/pypoman/}.

\bibitem{althoff2015introduction}
M.~Althoff, ``An introduction to {CORA},'' in \emph{W. App. Verif. Cont. \&
  Hyb. Syst.}, 2015, pp. 120--151.

\bibitem{MPT3}
M.~Herceg, M.~Kvasnica, C.~Jones, and M.~Morari, ``{Multi-Parametric Toolbox
  3.0},'' in \emph{Proc. Euro. Ctrl. Conf.}, 2013.

\bibitem{zonolab}
J.~Koeln, T.~J. Bird, J.~Siefert, J.~Ruths, H.~Pangborn, and N.~Jain,
  ``{zonoLAB: A MATLAB toolbox for set-based control systems analysis using
  hybrid zonotopes},'' \emph{arXiv preprint arXiv:2310.15426}, 2023.

\bibitem{boyd2004convex}
S.~Boyd and L.~Vandenberghe, \emph{Convex optimization}.\hskip 1em plus 0.5em
  minus 0.4em\relax Cambridge Univ. Press, 2004.

\bibitem{gleason2021lagrangian}
J.~Gleason, A.~Vinod, and M.~M. Oishi, ``Lagrangian approximations for
  stochastic reachability of a target tube,'' \emph{Automatica}, vol. 128,
  2021.

\bibitem{SReachTools}
A.~Vinod, J.~Gleason, and M.~M.~K. Oishi, ``{S}{R}each{T}ools: A {MATLAB}
  {S}tochastic {R}eachability {T}oolbox,'' in \emph{Proc. Hyb. Syst.: Comput.
  \& Ctrl.}, Montreal, Canada, 2019, \url{https://sreachtools.github.io}.

\bibitem{lipp2016variations}
T.~Lipp and S.~Boyd, ``Variations and extension of the convex--concave
  procedure,'' \emph{Opt. \& Engg.}, vol.~17, pp. 263--287, 2016.

\bibitem{gilbert}
I.~Kolmanovsky and E.~G. Gilbert, ``Theory and computation of disturbance
  invariant sets for discrete-time linear systems,'' \emph{Mathematical
  problems in engineering}, vol.~4, no.~4, 1998.

\end{thebibliography}
\end{document}